\documentclass[preprint,12pt,aps,prd]{revtex4-1}
\usepackage{amsmath}
\usepackage{graphicx}
\usepackage{epstopdf}
\usepackage{amssymb}
\usepackage{hyperref}
\usepackage{float}
\usepackage[T1]{fontenc}
\usepackage{tabularx,caption}
\usepackage{booktabs}
\usepackage{color}
\begin{document}
\def\be{\begin{equation}}
\def\ee{\end{equation}}
\def\ba{\begin{array}}
\def\ea{\end{array}}
\def\bc{\begin{center}}
\def\ec{\end{center}}
\newcommand{\ra}{\rangle}
\newcommand{\la}{\langle}
\newcommand{\eq}{\begin{eqnarray}}
\newcommand{\en}{\end{eqnarray}}
\newcommand{\bfq}{{\bf q}_{\perp}}
\newcommand{\bfql}{{\bf q}_{1\perp}}
\newcommand{\bfqr}{{\bf q}_{2\perp}}
\newcommand{\bfk}{{\bf k}_{\perp}}
\newcommand{\bfkpr}{{\bf k}_{\perp}^\prime}
\newcommand{\bfki}{{\bf k}_{\perp i}}
\newcommand{\bfb}{{\bf b}_{\perp}}
\newcommand{\bfbi}{{\bf b}_{\perp i}}
\newcommand{\bfe}{{\bf e}_{\perp}}
\newcommand{\bfP}{{\bf P}_{\perp}}
\newcommand{\bfPpr}{{\bf P}_{\perp}^\prime}
\newcommand{\bfp}{{\bf p}_{\perp}}
\newcommand{\bfn}{{\bf n}}
\newcommand{\bfS}{{\bf S}_{\perp}}
\newcommand{\bfppr}{{\bf p}_{\perp}^\prime}
\newcommand{\bfQ}{{\bf Q}_{\perp}}
\newcommand{\bfQpr}{{\bf Q}_{\perp}^\prime}
\newcommand{\bfD}{{\bf \Delta}_{\perp}}
\newcommand{\bfr}{{\bf r}_{\perp}}


\title{Generalized Parton Distributions of Pion for Non-Zero Skewness in AdS/QCD}
\author{Navdeep Kaur$^1$, Narinder Kumar$^2$, Chandan Mondal$^3$ and Harleen Dahiya$^1$}
\affiliation{$^1$Department of Physics, Dr. B. R. Ambedkar National Institute of Technology, Jalandhar-144011, India \\
$^2$Department of Physics, Indian Institute of Technology Kanpur, Kanpur-208016, India\\
$^3$Institute of Modern Physics, Chinese Academy of Sciences, Lanzhou-73000, China}
\begin{abstract}
We study the generalized parton  distributions (GPDs) of pion for non-zero skewness by considering the leading $| q \bar{q} \rangle$ Fock state component. Inspired from AdS/QCD light-front wave functions, we calculate the pion GPDs. By taking the Fourier transforms we obtain the results for impact-parameter dependent parton distribution functions (ipdpdf) and for GPDs in longitudinal boost-invariant space. We also calculate the charge density and gravitational form factor of the pion. The pion unpolarized transverse momentum distribution (TMD) have also been calculated. The results provide rich information on the internal structure of pion.
\end{abstract}
\maketitle

\section{Introduction}
A lot of progress has been made in the last two decades to understand the three-dimensional (3D) structure of hadrons. We are now facing a new era promising informative measurements on the structure of hadrons. In order to understand the 3D structure, light-front is the most suitable framework. Due to many exciting properties, this framework has been used in many theoretical models. In the recent years, exclusive scattering processes like deep virtual Compton scattering (DVCS) or deep virtual meson production (DVMP) prove to be an excellent way to probe the internal structure of hadrons. The hadron structure is encoded in the generalized parton distributions (GPDs).  GPDs help us to understand the 3D structure and they can be easily reduced to parton distribution functions, form factors, charge distributions, magnetization density and gravitational form factors  \cite{burk2000}. By taking a 3D Fourier transform of the electromagnetic form factors of hadron,  the information about the spatial distributions of charge (charge distribution) \cite{gerald} can be extracted. The second moment of charge distribution gives the mass  distribution (gravitational form factors (GFFs)) for hadron. It is very interesting to mention that the second  Mellin  moments  GPDs  give  the GFFs without actual gravitational scattering.
Several experiments, such as, H1 collaboration \cite{adloff}, ZEUS collaboration \cite{chekanov} and fixed target experiments at HERMES \cite{air} have finished taking data on DVCS. Experiments are also being done at JLAB, Hall A and B  and COMPASS at CERN to access GPDs of hadrons \cite{step}.

GPDs not only allow us to access partonic configurations with a given longitudinal momentum fraction (similar to deep inelastic scattering (DIS)), but also at specific (transverse) location inside the hadron. GPDs depend on three variables  $x$, $\zeta$ and $t$ where $x$ is the fraction of momentum carried by the active quark, $\zeta$ gives the longitudinal momentum transfer and $t$ is the square of the momentum transfer in the process. However, it has to be realized that only two of these variables (fully defined by detecting the scattered lepton = $x_b$, where $x_b$ is the Bjorken variable used in DIS) and $t$ (fully defined by detecting either the recoil proton or meson) are accessible experimentally.
Further, transverse momentum dependent parton distributions (TMDs) \cite{diehl} contain information on both the longitudinal momentum fraction and transverse momentum of partons in the hadron. They give a 3D view of the parton distribution in momentum space, complementary to what can be obtained through GPDs \cite{burk2000, diehl2002, ji2003, belitsky2004, boffi2007}. TMDs can be measured in a variety of reactions in lepton-proton and proton-proton collisions as semi-inclusive deep inelastic scattering (SIDIS) \cite{alessandro, ji} and Drell-Yan (DY) production \cite{baranov}.

Among the hadrons, pions are very fascinating particles and hold a lot of information on the structure of hadrons. From the DY process  with pion beams \cite{Drell:1970wh, Christenson:1970um} one can access the partonic structure of pion by hitting them on nuclear target \cite{Peng:2014hta, Chang:2013opa, Reimer:2007iy, McGaughey:1999mq}. Hadron structure is also influenced by chiral symmetry and its breaking. Chiral symmetry is dynamically broken in quantum chromodynamics (QCD) leading to generation of Goldstone bosons-pions-having small mass as compared to other hadrons. Pions are critical in providing the force that binds the protons and neutrons inside the nuclei and they also influence the properties of the isolated nucleons. Understanding of matter is not complete without getting a detailed information on the role of pions.  Therefore, it becomes important to expose the role played by the pions in understanding the hadron structure. GPDs of the pion have been discussed in various models like chiral quark model \cite{broniowski,dorokhov}, NJL model \cite{davidson}, double distributions \cite{polyakov}, light-front constituent quark model \cite{frederico} and lattice QCD \cite{brommel, dalley, dalley1}. In addition to this, GPDs of the nucleon have  also been studied by including the pion cloud contributions \cite{boffi2007, pasquini, pasquini1}. Pion-photon transition form factors have been widely discussed in Ref. \cite{lepage, tao, kroll, radyushkin1, hwang1, choi,xiao}. The Drell-Yan processes are the only source for the information on TMDs for hadrons other than nucleon \cite{badier, palestini, falciano, guanziroli, conway, bordalo}. Pion TMDs have been discussed in the light-front constituent approach \cite{Pasquini:2014ppa}.

The suitable wave functions for mesons are obtained in AdS/QCD model which relates the five dimensional AdS space to the Hamiltonian formulation of QCD on the light-front \cite{stan2004, stan2009, teramond, stan2008, stan2008:2}. To  study  the  3D  internal  structure of the pion, we will use light-front wave functions (LFWFs) from the soft-wall model of the AdS/QCD correspondence. At high energies, AdS/QCD will break down where QCD is asymptotically free but at low energy it gives reasonable results for the meson sector \cite{witten, babin, sakai, erlich, kruczen, rold, evans}.

In the present work, we present the results on the GPDs of the pion by considering the leading $| q \bar{q} \rangle$ Fock state contributions. GPDs are obtained from the overlap of  LFWFs inspired from the soft wall AdS/QCD predictions and we consider here the case when the skewness is non-zero. The situation $\zeta \ne 0$ has an important difference from $\zeta=0$ because as the pion loses the longitudinal momentum its transverse position is shifted by an amount which is proportional to $\zeta$. The LFWFs used here have several interesting properties which are consistent with data i.e., parton distributions, form factor and electromagnetic radii. The wave function contains two function $f(x)$ and $\bar{f}(x)$ having dependence on two parameters $x$ and $\zeta$. $f(x)$ is related with the pion pdf and $\bar{f}(x)$ represents the $Q^2$ dependence of the pion GPD and electromagnetic form factor. The wave functions $\psi^{(0)}(x, {\bf k}_\perp)$ and $\psi^{(1)}(x, {\bf k}_\perp)$ with $L_z=0$ and $L_z=1$ respectively give the correct scaling behavior and follow the quark counting rule \cite{thomas,gutsche}. Pion GPD is studied for both longitudinal and transverse position space by taking Fourier transform with respect to $\zeta$ and $\Delta_\perp$ giving the distribution of partons in the longitudinal and transverse position space  respectively. We further extend our calculations to get the results for charge density, gravitational form factors and for charge distributions in coordinate space. In the present work we also calculate the unpolarized TMD of the pion in AdS/QCD model.

The plan of paper is as follow. We define the pion GPDs in section II, ipdpdfs are defined in section III followed by the GPDs in longitudinal boost-invariant space in section IV. Charge density and gravitational form factors of the pion are  given in section V and VI respectively. In section VII we discuss the pion TMD. We summarize and and conclude our results in section VIII
.
\section{Generalized parton distributions of the pion}
In general, the spin non-flip GPD $H$ is defined through the off-forward matrix elements of the bilinear vector current as \cite{harind, harleen, LC-brodsky}
\begin{equation}
H(x,\zeta,t)=\int{\frac{dy^-}{8\pi}} e^{\iota x P^+ y^-/2} \langle P'|\bar{\psi}(0)\gamma^{+} \psi(y)|P\rangle|_{ y^+=0, \ y^\perp=0},
\end{equation}
where $ P $ and $ P^{'} $ are the momentum of pion with mass $ M $ in initial and final state, respectively. In the above expression, $ {\psi}(0) $ and $ {\psi}(y) $ are the quark fields at different points 0 and $ y $. The variable $ x $ is the light-front longitudinal momentum fraction carried by the struck quark, $ \zeta $ is the skewness parameter which measures the longitudinal momentum transfer and $ t = \Delta^2 $ is the square of four momentum transfer from the target.
In the AdS/QCD model, the pion with valence partons is considered as an composite system of a quark and antiquark. With the minimal Fock state configuration, i.e. $ q\bar{q} $, one can define the pion state with different values of $L_z=0,\pm 1$ as \cite{Burkardt:2002uc, Ji:2003yj, Pasquini:2014ppa} 
\eq
|\pi^+(P^+,\bfP)\ra &=& \ |\pi^+(P^+,\bfP)\ra_{L_z=0} +
|\pi^+(P^+,\bfP)\ra_{|L_z|=1} \,,\nonumber\\
|\pi^+(P^+,\bfP)\ra_{L_z=0}  &=&
\int \frac{d^2\bfk}{16\pi^3} \, \frac{dx}{\sqrt{6 x (1-x)}} \,
\psi_\pi^{(0)}(x,\bfk) \, \sum_{a=1}^{3} \,
\Big[b^{\dagger a}_{u\uparrow}(1) d^{\dagger a}_{d\downarrow}(2)
-    b^{\dagger a}_{u\downarrow}(1)
     d^{\dagger a}_{d\uparrow}(2)\Big] \, |0\ra \nonumber\\
|\pi^+(P^+,\bfP)\ra_{|L_z|=1}   &=&
\int \frac{d^2\bfk}{16\pi^3} \, \frac{dx}{\sqrt{2 x (1-x)}} \,
\psi_\pi^{(1)}(x,\bfk) \nonumber\\
&\times&  \, \Big(\frac{k_1 + ik_2}{\sqrt{3}} \, \sum_{a=1}^{3} \,
     b^{\dagger a}_{u\uparrow}(1)
     d^{\dagger a}_{d\uparrow}(2) \, |0\ra\,
+  \frac{k_1 - ik_2}{\sqrt{3}} \, \sum_{a=1}^{3} \,
    b^{\dagger a}_{u\downarrow}(1)
     d^{\dagger a}_{d\downarrow}(2)  \, |0\ra\,\Big),
\en
where $(1) = (xP^+,\bfk+x\bfP)$ and $(2) = \left( (1-x)P^+,-\bfk+(1-x)\bfP \right )$. Here $b^{\dagger a}_\lambda(b^a_\lambda)$ and $d^{\dagger a}_\lambda(d^a_\lambda)$ are the creation (annihilation) operators of $u$-quark and $\bar d$-quark, with color $a$, respectively. They obey the non vanishing anti-commutation relations
\eq
\{b^a_\lambda(k^+,\bfk),b^{\dagger a'}_{\lambda'}(k^{+'},\bfkpr)\} &=&
\{d^a_\lambda(k^+,\bfk),d^{\dagger a'}_{\lambda'}(k^{+'},\bfkpr)\}\nonumber\\
&=& 2 k^+ \, \delta_{\lambda\lambda'} \, \delta^{aa'} \,
\delta(k^+ - k^{+'}) \, \delta^2(\bfk-\bfkpr)\,.
\en
We consider the DGLAP region, with $\zeta < x < 1$, for the discussion of pion GPD which corresponds to the situation where one removes a quark from the initial pion with light-front longitudinal momentum $ (x + \zeta)P^{+} $ and re-insert it into the final pion with longitudinal momentum $ (x - \zeta)P^{+} $.


The spin non-flip GPD $ H_{\pi} $ in terms of the overlap of LFWFs is represented as
\eq
H_\pi(x,\zeta,t) &=&
\int\frac{d^2\bfk}{16\pi^3} \,
\biggl[ \psi^{(0)\dagger}_\pi(x^{'},\bfk') \,
     \psi^{(0)}_\pi(x,\bfk) \,+\,
     \bfk' \bfk \,
     \psi^{(1)\dagger}_\pi(x^{'},\bfk') \,
     \psi^{(1)}_\pi(x,\bfk)
\biggr], \label{hpi}
\en
where
\begin{equation}
 x^{'} =  \dfrac{x-\zeta}{1-\zeta}, \quad \quad  \bfk' = \bfk - \dfrac{1-x}{1-\zeta} \Delta_{\perp}.
\end{equation}
The LFWFs of pion, $  \psi^{L_{z}}_\pi(x,\bfk) $,  with total quark orbital angular momentum $L_{z} = 0,\pm1$ are given by~\cite{thomas,gutsche}
\eq
\psi_\pi^{(0)}(x,\bfk) &=& \frac{4\pi N_0}{\kappa} \,
\frac{\sqrt{\log(1/x)}}{1-x} \,
\sqrt{f(x) \, \bar f(x)} \, \exp\biggl[- \frac{\bfk^2}{2\kappa^2}
\frac{\log(1/x)}{(1-x)^2} \, \bar f(x) \biggr] \,, \nonumber\\
\psi_\pi^{(1)}(x,\bfk) &=& \frac{4\pi N_1}{\kappa^2} \,
\frac{\sqrt{\log^3(1/x)}}{(1-x)^2} \,
\sqrt{f(x) \, \bar f^3(x)} \, \exp\biggl[- \frac{\bfk^2}{2\kappa^2}
\frac{\log(1/x)}{(1-x)^2} \, \bar f(x) \biggr], \label{wf}
\en
where $ \kappa $ is the AdS/QCD scale parameter,  $N_0$ and $N_1$ are the normalization factors. The wave functions mentioned above are the generalizations of LFWFs of the pion in soft-wall AdS/QCD. The original pionic LFWF with $L_z=S_z=0$ \cite{meson-brodsky} has been extracted from light-front holography by considering the pion electromagnetic form factor in two approaches-AdS/QCD and light-front QCD. 
The wave functions adopted here (Eq. (\ref{wf})) are improved by introducing the profile functions $f(x)$ and $\bar{f}(x)$ \cite{thomas}
\eq
f(x)=x^{\alpha-1}(1-x)^\beta(1+\gamma x^\delta),\quad\quad \quad
\bar{f}(x)=x^{\bar{\alpha}}(1-x)^\beta(1+\bar{\gamma} x^{\bar{\delta}}),
\en
in the original pionic LFWF, where $\alpha, \bar{\alpha}, \beta, \gamma, \bar{\gamma}, \delta, \bar{\delta}$ are the free parameters. The improved LFWFs in Eq. (\ref{wf}) reduce to the original AdS/QCD LFWF in the limit $f(x)= \bar{f}(x)=1$.
The incorporation of the profile functions, $f(x)$ and $\bar{f}(x)$ in the LFWFs, has been made mainly to obtain the correct scaling behavior of pion PDF at large $x$ and the correct scaling of the pion electromagnetic form factor at large $Q^2$. The choice of the function $f(x)$ has been constrained by the pion PDF, $q_\pi(x) \sim f(x) $ \cite{aicher} whereas $\bar{f}(x)$ was fixed from the fit to correct $Q^2$ dependence of the pion electromagnetic form factors.
Both the functions have same scaling as $(1-x)^\beta$ at large $x$, which leads to correct $(\frac{1}{Q^2})^{\tau-1}$ (here $\tau$ is twist) power scaling of pion form factor at large $Q^2$ independent of the value of $\beta$ and this is consistent with quark counting rules.
Again, with the choice of $\beta=2.03$, the pion PDF at large $x$ is consistent with the modified E615 experimental data \cite{Conway:1989fs} reanalyzed by including the soft gluon resummation effects \cite{aicher}.
The values of the parameters $\alpha,~\beta,~\gamma$ and $\delta$ are taken from the Ref. \cite{aicher} which give the best fit for pion valence PDF, whereas the parameters $\bar{\alpha},~\bar{\gamma}$ and $\bar{\delta}$ are fixed by a fit to data on the electromagnetic form factor of the pion \cite{thomas}. Overall, the improved LFWFs are able to reproduce several fundamental properties of the pion consistent with data of valence parton distribution, electromagnetic form
factor and radius \cite{thomas}.

By substituting  Eq. (\ref{wf}) in Eq. (\ref{hpi}), we have the following expression for pion GPD $ H_\pi(x, \zeta ,t) $:
\begin{eqnarray}\label{gpd}
H_\pi(x, \zeta,t)&=& N_o^2 {\rm exp}\Bigg[\frac{1-B}{B}Q^2 {\rm log}\Bigg(\frac{1-\zeta}{x-\zeta}\Bigg) \frac{\bar{f}(x')}{2 \kappa^2}\Bigg] \times \nonumber\\
&&\Bigg[\frac{16 \pi^2 \sqrt{A} (1-\zeta)}{\kappa^2 B (1-x)^2} \Bigg(1+ \Bigg(\frac{N_1}{N_o}\Bigg)^2 \frac{A(1-\zeta)}{8 \pi \kappa^2 B^2 (1-x)^2}\Bigg)\Bigg] \times \nonumber\\
&&\Bigg[{\rm log}\Bigg(\frac{1-\zeta}{1-x}\Bigg)\Bigg(\frac{1-\zeta}{1-x}\Bigg)^2 \frac{\bar{f}(x')}{2 \kappa^2} \Bigg(1- {\rm log}\Bigg(\frac{1}{x}\Bigg) \frac{\bar{f}(x)}{2 \kappa^2 (1-x)^2}\Bigg)+ \nonumber\\
&&{\rm log}\Bigg(\frac{1}{x}\Bigg) \frac{\bar{f}(x)}{2 \kappa^2 (1-x)^2} \Bigg],
\label{gpdpion}
\end{eqnarray}
where
\eq
  A &=& \log\Bigg(\frac{1-\zeta}{1-x}\Bigg)\log\Bigg(\frac{1}{x}\Bigg) f(x) \bar{f}(x) f(x') \bar{f}(x'), \\
  B &=& \log\Bigg(\frac{1-\zeta}{1-x}\Bigg)\Bigg(\frac{1-\zeta}{1-x}\Bigg)^2 \frac{\bar{f}(x')}{2 \kappa^2}+ \log\Bigg(\frac{1}{x}\Bigg) \frac{\bar{f}(x)}{2 \kappa^2(1-x)^2},\\
  f(x)&=& \Bigg(\frac{x-\zeta}{1-\zeta}\Bigg)^{\alpha-1} \Bigg(\frac{1-x}{1-\zeta}\Bigg)^\beta \Bigg(1+ \gamma \Bigg(\frac{x-\zeta}{1-\zeta}\Bigg)^\delta\Bigg),\\
  \bar{f}(x)&=&  \Bigg(\frac{x-\zeta}{1-\zeta}\Bigg)^{\bar{\alpha}} \Bigg(\frac{1-x}{1-\zeta}\Bigg)^\beta \Bigg(1 + \bar{\gamma}\Bigg(\frac{x-\zeta}{1-\zeta}\Bigg)^{\bar{\delta}}\Bigg) ,
\en
with $ {\bf \Delta_\perp^2} = Q^2 = -(1-\zeta) t - \zeta^{2 } M^{2} $. 
For the numerical calculations, we have used the pion mass, $ M = 0.139$ GeV and rest of the parameters are taken from \cite{thomas}.

\begin{figure}
\begin{minipage}[h]{1.0\linewidth}
\centering
\small{(a)} \includegraphics[width=2.82 in]{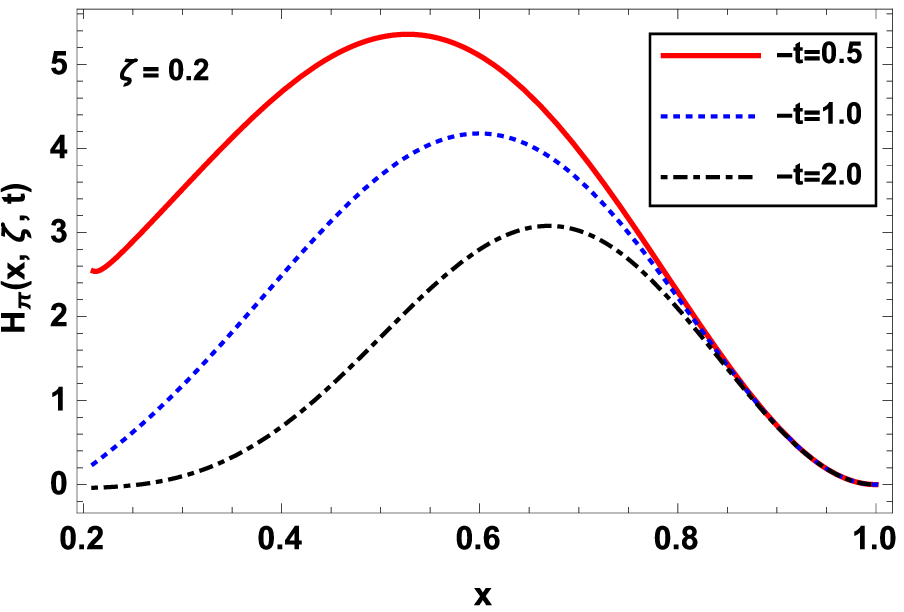}\hfill
\small{(b)} \includegraphics[width=2.8 in]{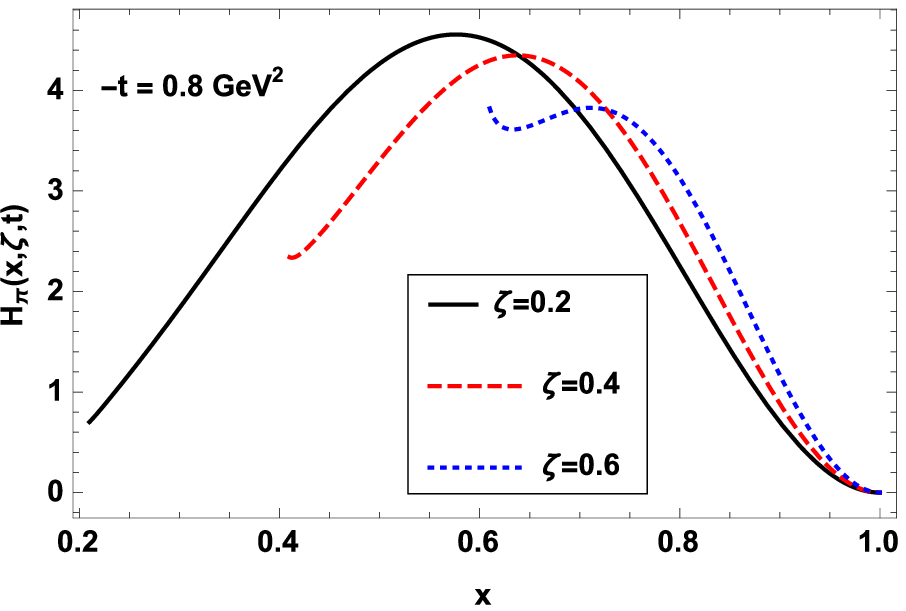}
\caption{The pion GPD $ H_\pi(x, \zeta, t)$ as a function of $x$ for (a)  different values of $-t$ (in GeV$^{2}$) and a fixed value of $\zeta=0.2$ and (b) different values of $\zeta$ and a fixed value of $-t=0.8 $  GeV$^{2}$.}
\label{Fig1}
 \end{minipage}
\end{figure}

\begin{figure}
\centering
\includegraphics[width=2.8 in]{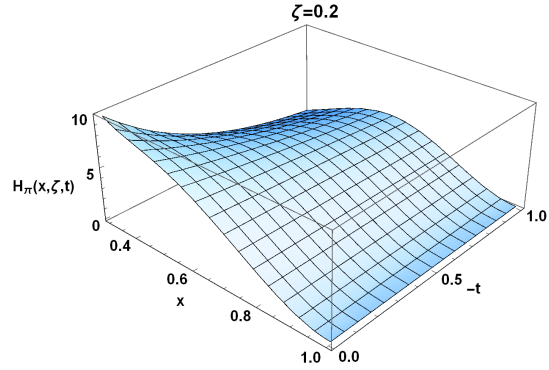}
\caption{3D plot of pion GPD $ H_\pi(x, \zeta, t)$ as a function of $x$ and $ -t $ (in GeV$^{2}$) for $\zeta =0.2 $.}
\label{Fig2}
\end{figure}

In most experiments, skewness is non-zero, so it is important to investigate the GPDs of pion with non-zero skewness. In Fig. \ref{Fig1} (a), we have presented the plots for pion GPD $ H_\pi(x, \zeta, t)$ as a function of $ x $ for the different values of $-t$ at the fixed value of $\zeta = 0.2$. We observe that the magnitude of distribution decreases with increase in the values of $-t$ and the peak shift towards the higher value of $x$. This suggests that as the momentum transfer during the process increases, the active quark carries more momentum. On the other hand, in Fig. \ref{Fig1} (b), we have shown the plots for pion GPD $ H_\pi(x, \zeta, t)$ as a function of $ x $ for the different values of $\zeta$ at the fixed the value of $-t = 0.8$ GeV$^{2}$. We observe that the magnitude also decreases with the increase in $ \zeta $ for a fixed value of $ -t $ and peaks shift towards the higher value of $x$ with increase in the value of $\zeta$. To obtain complete information on the pion GPD, we present in Fig. \ref{Fig2} the 3D plot of pion GPD $ H_\pi(x, \zeta, t)$ as a function of $ x $ and $ -t $ for the fixed value of $ \zeta = 0.2 $. A cursory look at Fig. \ref{Fig2} reveals that the peak is shifted towards the higher value of $ x $ with the increase in value of $ -t $. The parton distribution is maximum at lower values of four momentum transfer as well as when the longitudinal momentum carried by the struck quark is less. Similar calculations have also done in chiral quark model \cite{broniowski} and light front constituent quark model \cite{pace} but for the case of for zero skewness. The case pertaining to non-zero skewness has been reported for the first time here.

\section{Impact-parameter dependent parton distribution functions (IPDPDFs)}
In this section, we calculate the ipdpdf which describes the distribution of partons in transverse plane  providing spatial tomography of the hadron. The ipdpdfs are obtained by a two-dimensional Fourier transform of the non-flip GPD as \cite{ Diehl:2002he, burkardt2003, ralston}

\eq
{\cal H}_{\pi}(x, \zeta, \bfb) &=& \dfrac{1}{(2 \pi)^2}\int\mathrm{d^2 \bf{D}_{\bot}} e^{-i\bf{D}_{\bot} \cdot \bfb} H_\pi(x, \zeta, t)\nonumber\\
\en
where $ b = \vert\bfb\vert $ is the impact parameter. For non-zero skewness, the impact parameter $ \bfb $ is the Fourier conjugate to
$ \bf{D}_{\bot} $ = $P_{\bot}^{'}/(1 - \zeta) - P_{\bot} / (1+\zeta)$ = $\bfD / (1-\zeta^{2})$ where $ \bfD^2 =  -(1-\zeta) t - \zeta^{2 } M^{2} $.  In the case of $ \zeta \neq 0 $, the struck quark suffers a loss of longitudinal momentum proportional to $ \zeta $ due to the change in transverse position of partons in the initial and final pion state. In DGLAP domain $\zeta < x <1 $, the impact parameter $ b $ gives the location where the quark is pulled out and put back into the pion.
\begin{figure}
\centering
\includegraphics[width=2.8 in]{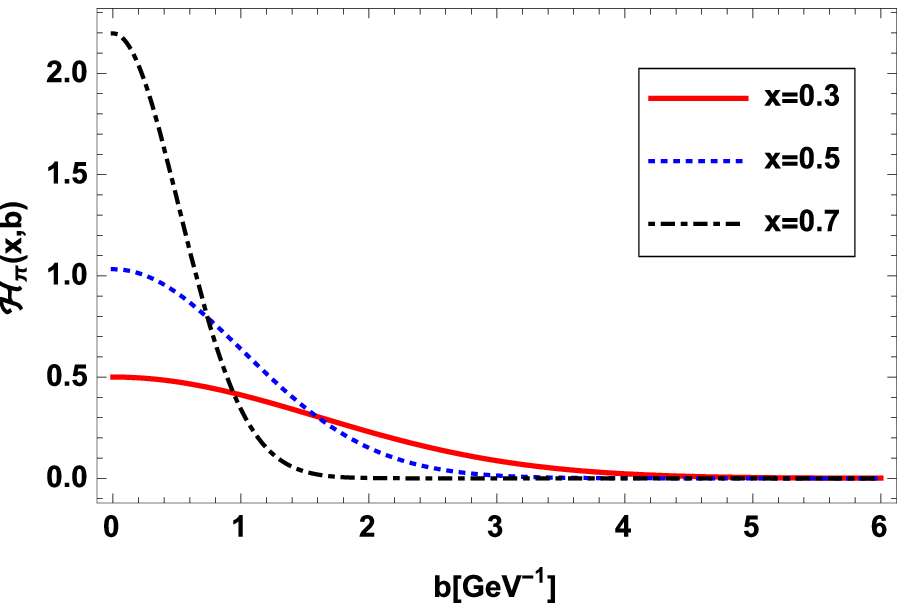}
\caption{Plot of ${\cal H}_\pi(x, b)$ against $ b = \vert \textbf{b} \vert$ with the different value of $ x $ for $ \zeta = 0 $.}
\label{fig3}
\end{figure}

In order to obtain the information about the distribution of quarks in transverse plane, we plot the ipdpdf in impact-parameter space. We show the plot of the ipdpdf ${\cal H}_\pi(x, b)$ against $ b $ with different values of $ x $ for zero skewness in Fig. \ref{fig3}. The zero skewness, $ \zeta = 0 $, implies that there is no longitudinal momentum transfer and hence probability interpretation is possible. We can see in plot that the magnitude of ${\cal H}_\pi(x, b)$ decreases as $ \vert b \vert $ increases which means that the distribution is more localized near the center of impact-parameter plane for the higher value of $ x $. This can interpreted as that the density of partons decreases as we move away from the center of transverse plane.
\begin{figure}
\centering
\small{(a)}\includegraphics[width=2.8 in]{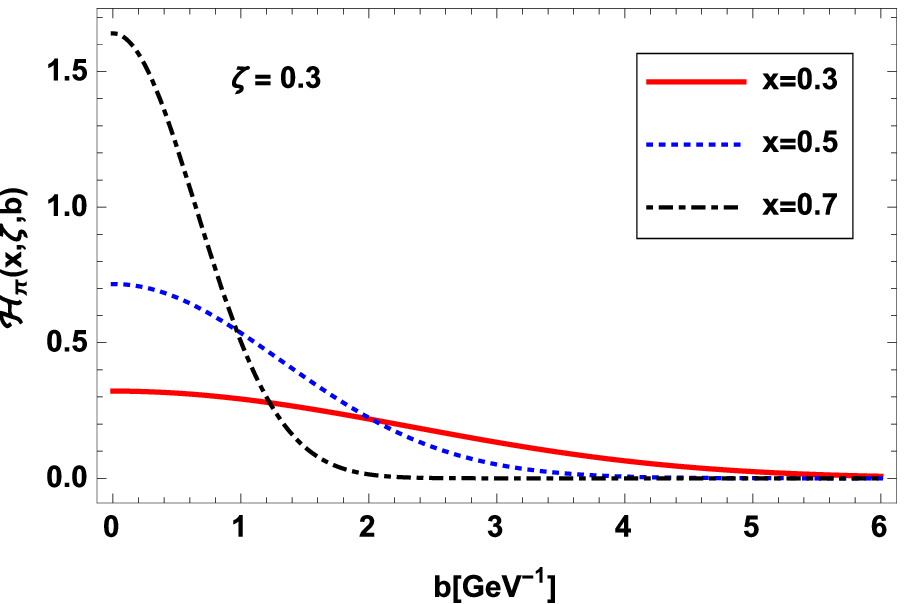}\hfill
\small{(b)}\includegraphics[width=2.8 in]{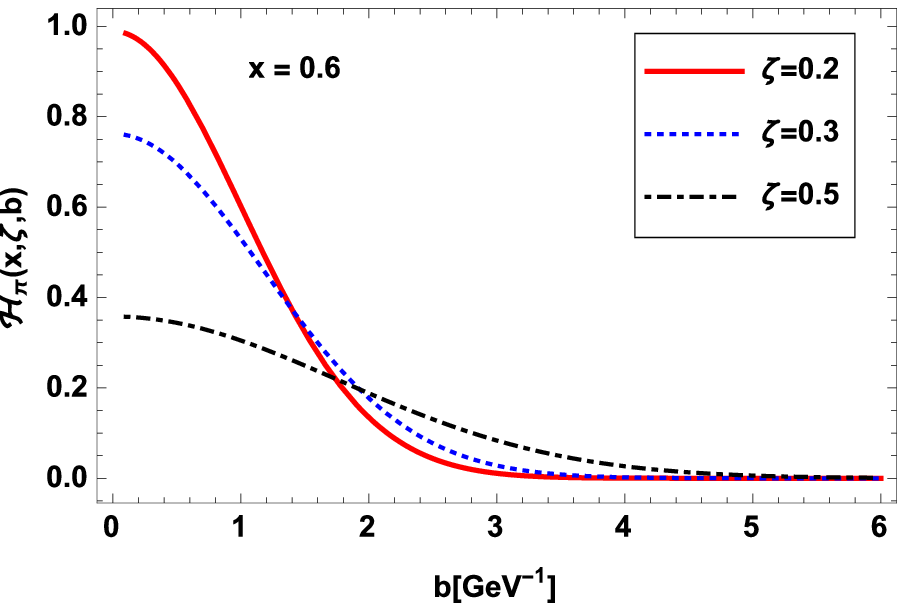}
\caption{Plots of ${\cal H}_\pi(x,\zeta ,b)$ against $ b $ with  (a) different values of $ x $ at a fixed value of $ \zeta=0.3 $ and (b)  different values of $ \zeta $ at a fixed value of $ x=0.6 $ .}
\label{fig4}
\end{figure}

We can also observe the dependence of ipdpdf on skewness to get more information on the distribution. The skewness dependent ipdpdf for non-zero skewness is shown in Fig. 4. In Fig. \ref{fig4} (a), we plot ${\cal H}_\pi(x,\zeta ,b)$  against $ b $ with  different values of $ x $ at a fixed value of $ \zeta = 0.2$. The behavior of ${\cal H}_\pi(x,\zeta ,b)$  in impact-parameter space is very similar to the behavior of ${\cal H}_\pi(x,b)$. The curve of ${\cal H}_\pi(x, \zeta,  b)$ becomes more and more narrow as we increase the value of $x$ which reflects that the most of the partons with large longitudinal momentum fraction are found near the center of impact-parameter space.
For a better understanding of  the skewness dependence of ipdpdf  we plot  ${\cal H}_\pi(x,\zeta ,b)$ against $ b $ with different values of $ \zeta $ but at a fixed $ x=0.6 $ in Fig. \ref{fig4} (b). We can clearly see that the distribution curve of ${\cal H}_\pi(x, \zeta, b)$  is narrower at the low value of $\zeta$ and become more and more wider with increase in the value of $\zeta$. This implies that in impact-parameter plane, a large number of partons are concentrated near the center of momentum at small longitudinal momentum transfer and as the longitudinal momentum transfer increases, the partons start spreading in impact-parameter space. The peak of ${\cal H}_\pi(x, \zeta,  b)$ is shifted downward with increase in the longitudinal momentum transfer. The calculations of ipdpdfs have been carried in the chiral quark models \cite{broniowski}, in transverse lattice formalism \cite{dalley} and in two component spectator model \cite{luiti} for zero skewness.

\section{Pion GPD in longitudinal boost-invariant space}
The Fourier transform of GPDs with respect to the skewness variable $ \zeta $ gives the GPDs in longitudinal boost-invariant coordinate space (longitudinal impact-parameter space). The skewness variable $ \zeta $  conjugate to the longitudinal boost invariant impact parameter is defined as $ \sigma = \dfrac{1}{2} b^{-}P^{+} $. The DVCS amplitude of a dressed electron in a QED model shows an interesting diffraction pattern in the longitudinal impact-parameter space  \cite{harind}. This is analogous to the diffraction scattering of a wave in optics where the distribution in $\sigma$ measures the physical size of the scattering  system in one-dimensional system.  In different phenomenological models, hadron GPDs show a similar behavior in longitudinal boost-invariant space \cite{harind, manohar}. We have observed a similar behavior for the pion GPD in longitudinal boost-invariant space in the AdS/QCD model. The expression for GPD in longitudinal boost-invariant space is
\eq
 {\cal H}_\pi(x,\sigma , t) &=& \frac{1}{2\pi}\int_{0}^{\zeta_{f}}\,\mathrm{d}\zeta \,\exp[i \sigma \zeta]\, H_\pi(x, \zeta , t),
 \en 	
 where $ \zeta_{f} $ acts as slit width. For the occurrence of diffraction pattern, a finite slit width provides a necessary condition. As we are considering DGLAP region, if $ x > \zeta_{max} $, then the upper limit of $ \zeta $ integration $ \zeta_{f} $ is given by $\zeta_{max} $ and if $ x < \zeta_{max} $ then it is given by $ x $. For a fixed value of $ -t $, the maximum value of $ \zeta $  is given as \cite{mondal, Mondal:2017wbf, chakrabarti}
\eq
\zeta_{max} &=& \frac{-t}{2 M^{2}}\Bigg(\sqrt{1+\frac{4 M^{2}}{(-t)}} -1 \Bigg).
\en
 \begin{figure}
\centering
\includegraphics[width=2.8 in]{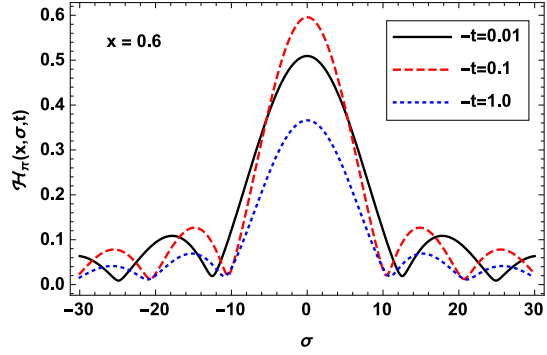}
\caption{Pion GPD ${\cal H}_\pi(x, \sigma ,t)$ in longitudinal boost-invariant space with the different values of $ -t $ (in  GeV$^2 $) for $ x $ = 0.6.}
  \label{fig5}
  \end{figure}
To obtain the information about the longitudinal size of the partons distributions, we plot the pion GPD $ {\cal H}_\pi(x,\sigma , t) $ in longitudinal boost-invariant space with different values of $ -t $ (in  GeV$^2 $) for fixed value of $ x $ = 0.6 in Fig.  \ref{fig5}. We observe a diffraction pattern for GPD in longitudinal boost-invariant space where $\zeta_{max}$ plays the role of slit width. The position of minima which is measured from the center of the diffraction pattern is inversely proportional to slit width and minima  moves away from the center as $\zeta_{max}$ decreases.  The distribution has primary maxima at $ \sigma = 0 $ followed by a series of secondary maxima. With the increase in $ -t $, the curve become narrower and the minima shifts toward the lower value of $ \sigma $. It reflects that the longitudinal size of the distribution of partons become longer and conjugate shape of light-cone momentum distribution becomes narrower. Further, position of minima is always independent of the helicity which is the characteristic of diffraction obtained in the single slit experiment. The nucleon GPDs have been studied in longitudinal boost-invariant space for non-zero skewness \cite{harind,  manohar, mondal, chakrabarti, harindra, chandan, kumar,Mondal:2017wbf} but pion GPD in the longitudinal boost-invariant space has not been attempted so far.

\section{Charge density of pion}
In this section, we discuss the pion charge density in AdS/QCD model which is the matrix element of the light-front density operator integrated over longitudinal distance. It gives the probability that the charge located at a transverse distance $ b $ from the transverse center of momentum irrespective of the value of the longitudinal position or momentum \cite{miller}.  The pion GPD for zero skewness, $ H_{\pi}(x, \zeta = 0, t) $, is related to the  pion electromagnetic form factor as \cite{thomas, broniowski}
\eq
F_{\pi}(-t= \bfD^2) &=&  \int_{0}^{1} dx \, H_{\pi}(x, \zeta = 0, t).
\en
Pion GPD with zero skewness is already defined in Ref. \cite{thomas}  and one can also obtain it by setting $\zeta=0$ in Eq. (\ref{gpdpion}). The charge density of the pion in the transverse plane as \cite{gerald}
\eq
\rho_{\pi}(\bfb)
&=& \frac{1}{4\pi^2} \int d^2\bfD \,F_\pi(\bfD^2)\, e^{-i\bfb\cdot\bfD}\nonumber\\
 &=& \dfrac{1}{2\pi}\int \mathrm{d\Delta}\, J_{0}(\Delta \ b) \, F_\pi(\bfD^2) ,
\en
where $\Delta=|\bfD|$ and $b=|\bfb|$. On the other side, the charge distribution in transverse coordinate space is defined as \cite{kim,Mondal:2016xsm}
 \eq
P_{\pi}(\bfr) &=&  \int dx \, P_{\pi}(x,\bfr)\nonumber\\
 &=& \int dx \, \big[ \tilde{\psi}^{(0)\dagger}_\pi(x,\bfr) \,
     \tilde{\psi}^{(0)}_\pi(x,\bfr) \,+\,
     \tilde{\psi}^{(1)\dagger}_\pi(x,\bfr) \,
     \tilde{\psi}^{(1)}_\pi(x,\bfr)
\big] .
\en
The two dimensional Fourier transform of the LFWFs in momentum space $ \psi_{\pi}(x,\bfk) $ gives the LFWFs in transverse coordinate space  $\tilde{\psi}_{\pi}(x,\bfr) $ as
\eq
\tilde{\psi}_{\pi}(x,\bfr)
&=& \frac{1}{4\pi^2} \int d^2\bfk\, \psi_{\pi}(x, \bfk)\,e^{i\bfk\cdot\bfr},
\en
where $r = \vert\bfr\vert $ measures the distance between the active quark and spectator system. Both $ r $ and $ \bfb $ are conjugate to the momentum $ k $ and $ \bfD $ respectively, but still the charge density in impact-parameter space $ \rho_\pi(b) $ and in coordinate space $P_\pi(r)$ are not same.
%
\begin{figure}
\centering
\small{(a)}\includegraphics[width=2.8 in]{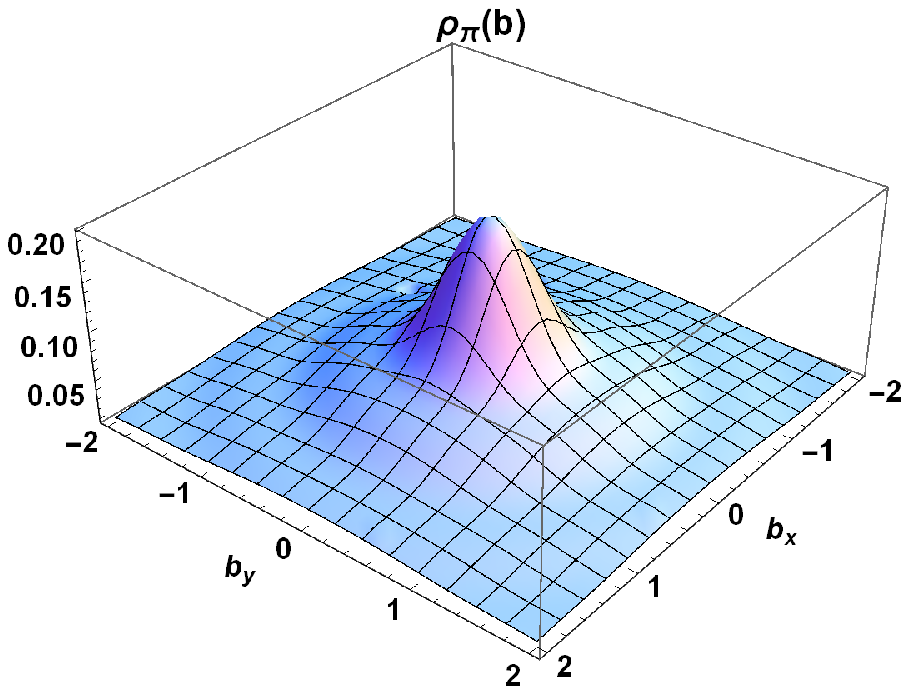}
\small{(b)}\includegraphics[width=2.8 in]{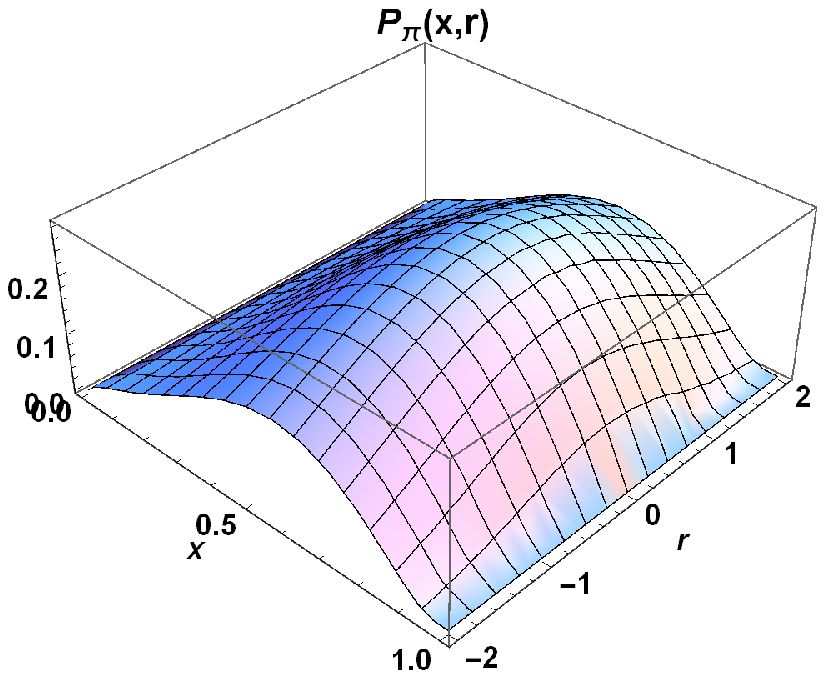}
\caption{The 3D plots of (a) $ \rho_\pi(b) $ and (b) $P_\pi(x,r)$ for pion in both transverse coordinate space and impact-parameter space.}
\label{fig:)Longitudinal boost invariant coordinate}
\label{fig7}
\end{figure}

To observe the charge distribution of valence quark inside pion in impact-parameter space and in transverse coordinate space, we calculate the charge distribution for both the cases.
In Fig. \ref{fig7} (a), we present the 3D plots of charge distribution $ \rho_\pi(b) $ in transverse plane. We observe that the charge distribution is axially symmetric in transverse plane. For a  more detailed information we present the 3D plot of longitudinal momentum distribution $P_\pi(x, r)$ for pion as a function of $ x $ and $ r $ is shown in Fig. \ref{fig7} (b). The distribution $P_\pi(x, r)$ shows the peak of distribution around the center $ x = 0.6 $ and near  $r=0$ but it decreases as the value of $r$ increases.
By comparing the two plots, we observe that the magnitude of  $ \rho_\pi(b) $ in impact-parameter space is slightly larger than the magnitude of  $P_\pi(r)$  in transverse coordinate space. It can be clearly seen in plots that both $\rho_\pi(b)$ and $P_\pi(r)$ have peaks at the low value of $ b$ and  $r $ space respectively, which implies that the large number of partons are concentrated near the center of plane. One can also notice that  width of the distributions in transverse coordinate space is larger than that in the impact-parameter space.  We also observe that $\rho_\pi(b)$ falls off faster than $P_\pi(r)$. Pion transverse charge distribution has been studied in \cite{miller, miller2009, marco, bing, dong}.

\section{Gravitational form factor of the pion}
Gravity plays an important role at both cosmic and Planck scale. In the subatomic level, the effect of gravity is absent but it is interesting to see that gravitational form factors can be obtained from the GPDs without actual gravitational scattering \cite{hwang}.  Gravitational form factors have interesting interpretations in impact-parameter space as discussed in \cite{abidin, selvugin}. In the present work, we have discussed one of the gravitational form factor $ A_\pi(Q^2) $ which gives the momentum fraction carried by each constituent of a pion and can be defined in terms of the overlap of LFWFs. For the quark in the pion we have
\eq
A_\pi^{q}(Q^2) &=& \int_{0}^{1} dx \, x \, H(x, \zeta = 0, t = - Q^2)\nonumber\\
&=&
\int\frac{d^2\bfk dx}{16\pi^3} \,x
\biggl[ \psi^{(0)\dagger}_\pi(x^{'},\bfk') \,
     \psi^{(0)}_\pi(x,\bfk) \,+\,
     \bfk' \bfk \,
     \psi^{(1)\dagger}_\pi(x^{'},\bfk') \,
     \psi^{(1)}_\pi(x,\bfk)
\biggr], \label{quark}
\en
and for the antiquark we have
\eq
A_\pi^{\bar{q}}(Q^2) &=&
\int\frac{d^2\bfk dx}{16\pi^3} \,(1-x)
\biggl[ \psi^{(0)\dagger}_\pi(x^{'},\bfk') \,
     \psi^{(0)}_\pi(x,\bfk) \,+\,
     \bfk' \bfk \,
     \psi^{(1)\dagger}_\pi(x^{'},\bfk') \,
     \psi^{(1)}_\pi(x,\bfk)
\biggr].\nonumber \label{antiquark}
\\
\en
Gravitational form factor is related to the second moment of GPD as can be seen from Eq. (\ref{quark}).
 \begin{figure}
\centering
\small{(a)}\includegraphics[width=2.5 in]{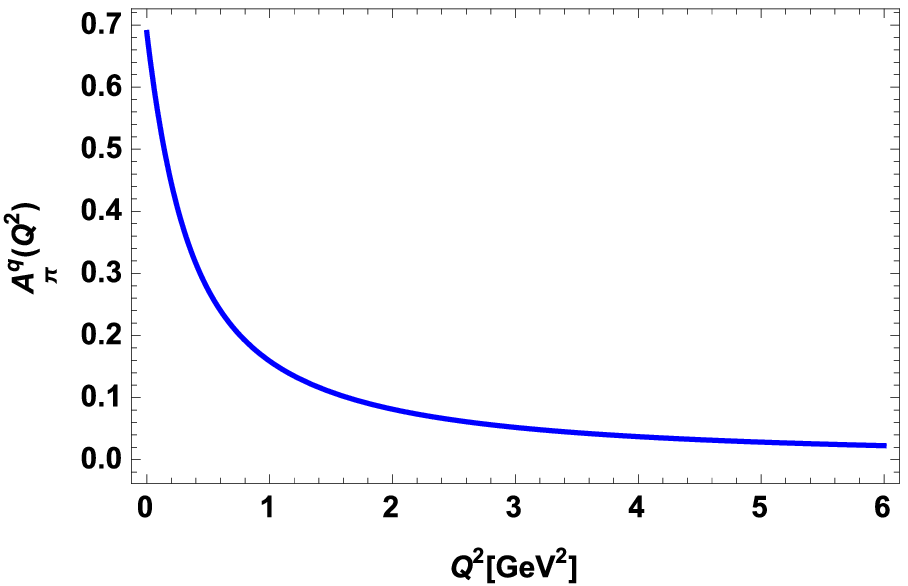}\hfill
\small{(b)}\includegraphics[width=2.5 in]{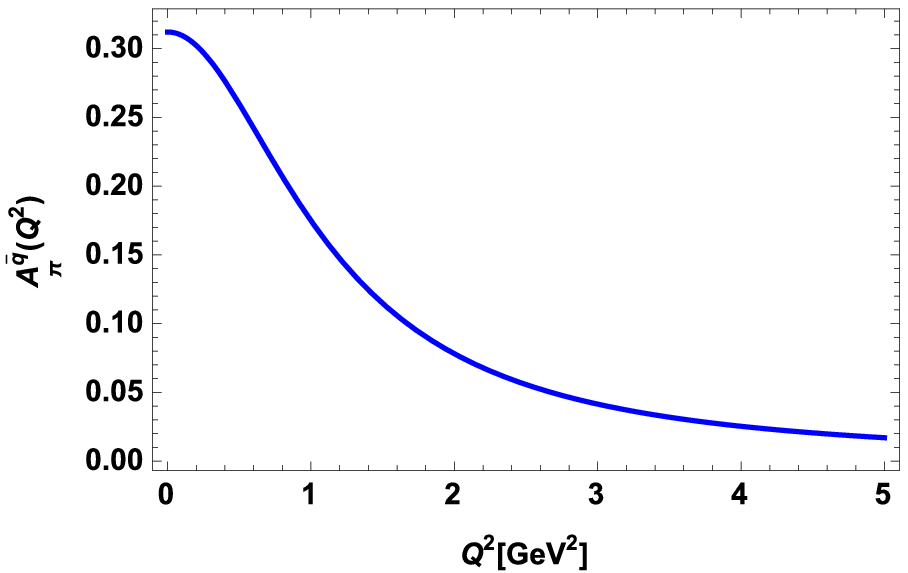}
\caption{Gravitational form factor as a function of $ Q^2 $ (a) for quark $ A_{\pi}^{q}(Q^2) $ and (b) for antiquark $ A_{\pi}^{\bar q}(Q^2) $.
\label{fig:)Gravitational form factor for quark and diquark}}
\label{fig8}
\end{figure}
Using the gravitational form factor of quark and antiquark for pion calculated from Eqs. (\ref{quark}) and (\ref{antiquark}), in Fig. \ref{fig8}, we present the results of gravitational form factor for quark $ A_{\pi}^{q}(Q^2) $ and antiquark $ A_{\pi}^{\bar q}(Q^2) $ as a function of $ Q^2 $. We observe that $ A_{\pi}^{q}(Q^2) $ and $ A_{\pi}^{\bar q}(Q^2) $ both decrease with the increase in $ Q^2 $ where the decrease in the case of $ A_{\pi}^{q}(Q^2) $ is much steep. The values of gravitational form factor $ A_{\pi}^{q}(Q^2) $ and $ A_{\pi}^{\bar q}(Q^2) $ at $Q^2=0$ are given in Table 1. One notices that at zero momentum transfer, the gravitational form factor satisfies the sum rule $ A_{\pi}(0) = A^q_{\pi}(0) + A^{\bar{q}}_{\pi}(0) \simeq 1.00$.
 \begin{figure}
\centering
\includegraphics[width=2.5 in]{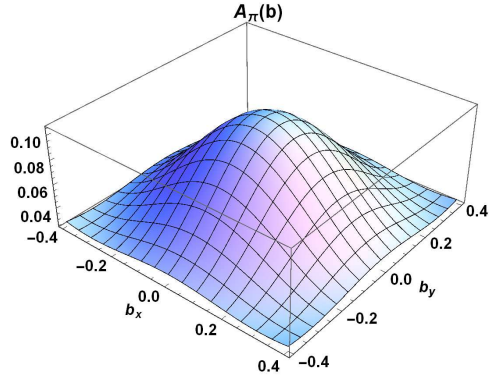}
\caption{Longitudinal Momentum density of pion $ A_{\pi}(b) $ in the transverse plane.
\label{fig:)Gravitational form factor}}
\label{fig9}
\end{figure}
\begin{table}[ht]
\caption{Gravitational form factor at $Q^2=0$.}
\centering
\begin{tabular}{c c}
\hline \hline        
 $A^q(0)$   \hspace{0.8 cm} & \vline \hspace{0.8 cm} 0.6878 \\
\hline
 $A^{\bar{q}}(0)$ \hspace{0.8 cm} & \vline \hspace{0.8 cm} 0.3121 \\ [1ex]
\hline     \hline
\end{tabular}
\end{table}

To get a more detailed information about the matter (i.e. gravitational charge) distribution in the pion, we study the longitudinal momentum density of pion. The longitudinal momentum density in transverse impact-parameter space in AdS/QCD model can be calculated by taking the Fourier transfer of the gravitational form factor \cite{abidin, selvugin, Mondal:2016xsm, Mondal:2015fok, Chakrabarti:2015lba, Kumar:2017dbf} and we have
\eq
A_{\pi}(b)
&=& \frac{1}{2\pi} \int_{0}^{\infty} dQ \, Q \, J_{0}(b Q) \,A_{\pi}(Q^2)\,,
\en
We present the plot for the longitudinal momentum density of pion $ A(b) $ in transverse plane as a function of impact-parameter space in Fig. \ref{fig9}. One can observe that the momentum density is axially symmetric and has the peak at the center ($ b = 0 $). By comparing the longitudinal momentum density in transverse plane with the charge density in transverse plane (shown in Fig. \ref{fig7} (a)), we conclude that the charge density spreads out more than the momentum density. This implies that the momentum density in the transverse plane is more compact than the charge density in the transverse plane.

\section{Pion transverse momentum distribution}
The TMDs of pion can be defined by using the correlator given below \cite{barbara, Meibner, boer}
\eq
\label{tmd=eqn:22}
\Phi(x,\bfk;S) &=& \frac{1}{2}\int\frac{dy^{-}d^{2}\textbf{\textit{y}}_{\perp}}{16\pi^3} e^{i\textit{k}.y} \langle P,S\vert \bar{\psi}(0)\, {\cal W} _{[0,\textit{y}]}\,\psi(\textit{y}) \vert P,S\rangle \Big\vert_{{\textit{y}}^{+}=0},
\en
where $ {\cal W}_{[0,\textit{y}]} $ is called gauge link operator or Wilson line which connects the quark field $ \psi $ at different points $0$ and $ \textit{y} $. It is taken to be unity in the present case. We calculate the unpolarized pion TMD at leading twist obtained from the above correlator. The spin factor in the expression for distribution functions for the pion is zero. Only spin independent function (i.e. unpolarized quark distribution) will remain. One can write the unpolarized pion TMD $f_1^\pi(x,\bfk^2)$ as an overlap of LFWFs:
\begin{eqnarray}
f_{1}^{\pi}(x,\bfk^{2})&=& \dfrac{1}{(2 \pi)^3}\big[ \vert \psi^{(0)}_\pi(x,\bfk)\vert^{2}+\bfk^{2} \, \vert\psi^{(1)}_\pi(x,\bfk)\vert^{2}
\big],\nonumber\\
&=& \frac{1}{\pi \kappa^2} \frac{\log(1/x)}{(1-x)^2} f(x) \bar{f}(x) \ \exp\bigg[-\frac{k^2 \log(1/x)}{\kappa^2 (1-x)^2}\bar{f}(x)\bigg] \times \nonumber\\
&& \Bigg(1+\frac{k^2}{\kappa^2} \Bigg(\frac{N_1}{N_0}\Bigg)^2 \Bigg(\frac{\log(1/x)}{1-x}\Bigg)^2 \bar{f}(x)^2\Bigg).
\end{eqnarray}
The unpolarized distribution function $ f_{1}^{\pi}(x,\textit{\textbf{k}}_{\bot}^{2}) $ describes the probability of finding a quark with longitudinal momentum fraction $ x $ and transverse momentum $ \bfk $ of the pion.
 \begin{figure}
\centering
\small{(a)}\includegraphics[width=2.5 in]{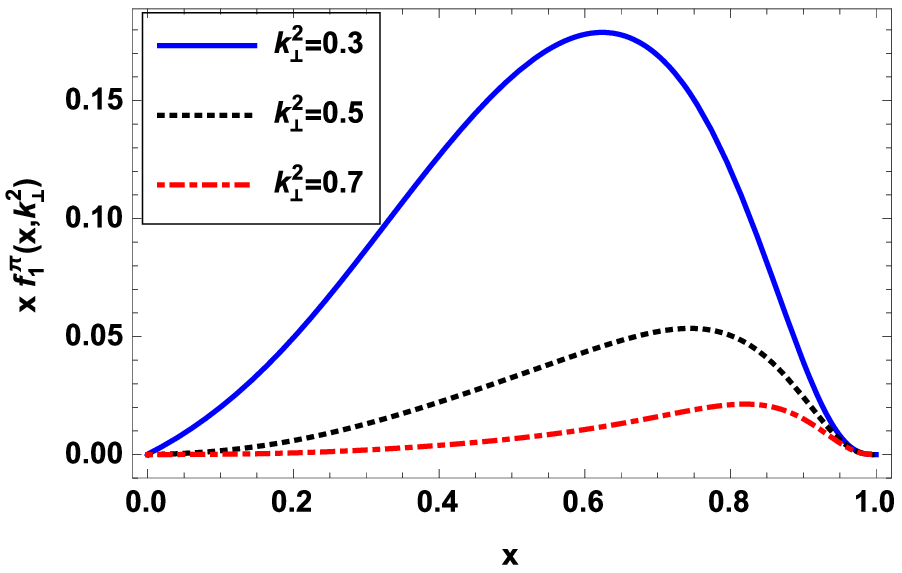}\hfill
\small{(b)}\includegraphics[width=2.5 in]{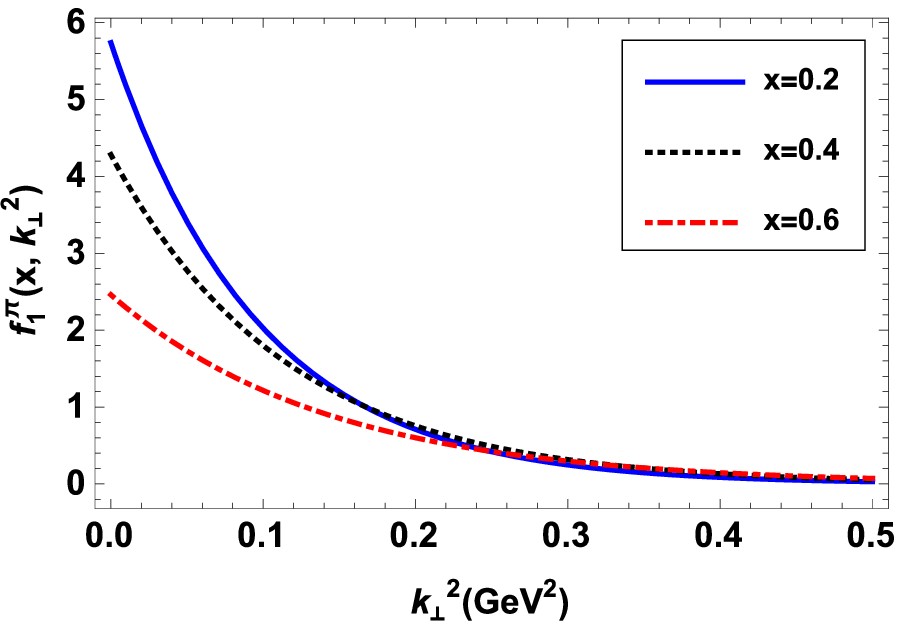}
\caption{Plots of pion unpolarized TMD $ x f_{1}^{\pi}(x, \bfk^2)$ (a) as a function of $x$ for the different values of $ \bfk^2 $ (in GeV$^2$) (b) as a function of $\bfk^2$ for the different values of $ x $.}
\label{pion-tmd}
\end{figure}
 \begin{figure}
\includegraphics[width=2.5in]{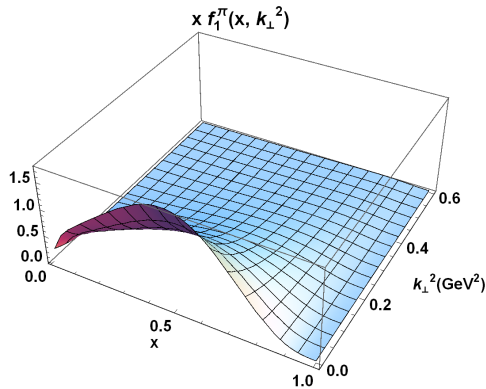}
\caption{3D plot of pion unpolarized TMD as a function of $ x $ and $ \bfk^2 $.}
\label{3d-pion-tmd}
\end{figure}
We show the plots of pion unpolarized TMD $ x f_{1}^{\pi}(x, \bfk^2)$ as a function of $ x $ for the different values of $ \bfk^2 $ in Fig. \ref{pion-tmd} (a) and as a function of $ \bfk^2 $ for the different values of $ x $ in Fig. \ref{pion-tmd} (b). We can see in Fig. \ref{pion-tmd} (a) that the peak of distribution shifts toward higher value of $ x $ with the increase in the square of transverse momentum $ \bfk^2 $.
In Fig. \ref{pion-tmd} (b), the dependence of $ x f_{1}^{\pi}(x, \bfk^2)$ on $ \bfk^2 $ is shown. The amplitude of $ x f_{1}^{\pi}(x, \bfk^2)$ decreases with the increase in transverse momentum $ \bfk^2 $ for a fixed value of $ x $.
We also shown the 3D plot of pion unpolarized TMD as a function of $ x $ and $ \bfk^2 $ in Fig. \ref{3d-pion-tmd}. Here the distribution peak is around $x=0.45$.

\section{Summary and Conclusions}
In the present work, we have studied the pion GPD $H(x, \zeta, t)$ with non-zero skewness in soft-wall AdS/QCD model. We have calculated the spin non-flip GPD in terms of the overlaps of light-front wave functions in the DGLAP region (for $ x > \zeta $).
The results are shown for ipdpdf for zero skewness as well as for non-zero skewness. In the model, we have observed a similar behavior for the ipdpdf in both the cases. The distribution is more localized near the center of momentum at large longitudinal momentum fraction. The ipdpdf gives complete information on the spatial distribution of partons inside pion.

We have also presented the results of pion GPD in longitudinal boost invariant space $\sigma$. We have observed a diffraction pattern for the pion GPD in boost invariant space in this model as observed for the nucleon GPDs in the different  phenomenological models. The diffraction pattern is observed for small value of $ -t $ and a dip appears at the center (at $ \sigma = 0 $). Further, the charge distribution for the pion in transverse coordinate space and in impact-parameter space have been calculated and it has been found that charge distribution for the pion in transverse coordinate space decreases more slowly as compared to  the charge distribution for the pion in  impact-parameter space.

Furthermore, we have shown the dependence of gravitational form factor for quark and antiquark on $ Q^2 $ in the model. The gravitational form factor satisfies the momentum sum rule. We have evaluated the longitudinal momentum density of pion in impact-parameter space and it comes out to be symmetric in nature. The momentum density in the transverse plane is more compact than the charge density. Finally, the pion unpolarized TMD in AdS/QCD model has been discussed to get a complete picture of the structure of pion. The dependence of TMD on $ \bfk^2 $ has also been shown.
\section{Acknowledgements}
N.K. acknowledges financial support received from Science and Engineering Research Board a statutory board under Department of Science and Technology, Government of India (Grant No. PDF/2016/000722, Project-"Study of Three Dimensional Structure of the Nucleon in Light Front QCD") under National Post-Doctoral Fellowship. This work of C.M. is supported by the funding from the China Post-Doctoral Science Foundation under the Grant No. 2017M623279.

\end{document}